# Two-step distortion-free reconstruction scheme for holographic microscopy

F. Joud[1], F. Verpillat[1], M. Lesaffre[2], N. Verrier[3], M. Gross[*3]

[1]*Laboratoire Kastler Brossel - UMR 8552 CNRS-ENS- Université Pierre et Marie Curie 24, rue Lhomond 75231 Paris cedex 05;* [2]*Institut Langevin - UMR 7587 CNRS-ESPCI Paristech 1, rue Jussieu 75005 Paris ;* [3]*Laboratoire Charles Coulomb - UMR 5221 CNRS-UM2 Université Montpellier II Place Eugène Bataillon 34095 Montpellier cedex*
*\*gross@lkb.ens.fr*

**Abstract:** We propose a three-dimensional holographic reconstruction procedure applicable with no *a priori* knowledge about the recording conditions enabling distortion-free three-dimensional object reconstruction. This scheme is illustrated through reconstruction of gold nano beads in Brownian motion.

## 1. Introduction

Hologram reconstruction most often consists in back-propagating the hologram from the sensor plane to the object plane [1-3]. This classical procedure is denote by the "1" arrow of the experimental set-up depicted Fig. 1. This method is not suitable for distortion-free reconstruction of a three-dimensional object. As a matter of fact, the imaging system magnification OM'/OM depends on the position of the M point of the three-dimensional object. For instance, if the object is located either side of the focal plane F, image of the object will exhibit strong distortions due to the fact that some image of the object points will be positioned at the infinity.

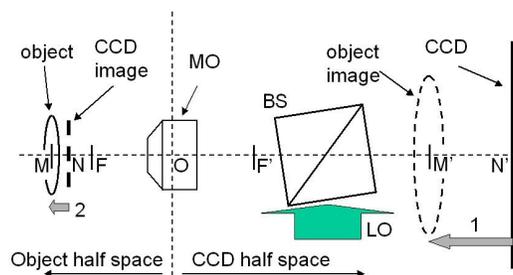

**Fig. 1** – Holographic microscopy setup. CCD: CCD sensor; BS: beam splitter; LO: Reference wave (Local Oscillator); MO: Microscope Objective; O: MO optical center; F and F': object and image focal points of MO. M: object position; M': image of the object position.

Instead of performing reconstruction in the CCD half-space, we propose to perform holographic reconstruction in the object half-space. Object field is therefore reconstructed by back-propagating the hologram from the CCD image plane to the object plane (see "2" arrow on Fig. 1). Thus, reconstruction is realized in two-steps: first, the hologram is propagated to the CCD image plane, and then propagation is calculated from the CCD image plane to the object plane.

## 2. Calculation of the corrected hologram in the CCD image plane

Digital hologram allow to access to the optical field in the CCD plane, which is the image, through MO, of the field in the CCD image plane. From the hologram, assessment of the field in the CCD plane is possible by

(i) Compensating the off-axis tilt introduced between reference and object beams, and performing a spatial filtering in Fourier space to keep only information of the +1 diffraction order [4]
(ii) Correcting the sphericity of the reference field.

To obtain the field in the CCD image plane, one also have to

(iii) Compensate for phase curvature induced by the MO
(iv) Modify CCD pixel pitches considering the CCD image magnification factor G=ON/ON'.

For off-axis compensation and optimal filtering, the image of the MO pupil $\tilde{H}(k_x, k_y, z)$ is reconstructed, from the hologram $H(x,y,0)$ using a single FFT formalism [5, 6]

$$\tilde{H}(k_x, k_y, z) = \mathsf{F}\left\{H(x,y,0)\exp\left[ik(x^2+y^2)/2z\right]\right\}$$

with $\mathsf{F}$s denoting the FFT operator, and $k = 2\pi/\lambda$ is the modulus of the wave vector. Coordinates in direct and Fourier spaces are respectively given by $(x,y)$, and $(k_x, k_y)$.

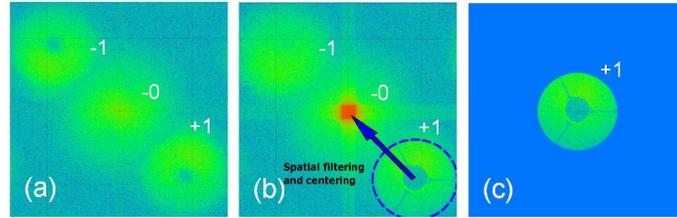

**Fig. 2** – Log scaled, false colored images of (a) Fourier transform of the hologram. (b) MO pupil image reconstructed using a single FFT algorithm. (c) Off-axis compensation and optimal filtering in Fourier space.

These aspects are illustrated by Fig. 2. Figure 2(a) is the Fourier transform of the acquired hologram. The +1, 0, and -1 off-axis induced diffraction orders are clearly visible. Calculating light back-propagation according to Eq. , allows to bring the MO pupil to focus (Fig. 2(b)). As far as all the relevant information is contained in the MO pupil, it is possible to perform an optimal spatial filtering around the +1 diffraction order. Finally, the +1 is brought to the center of the Fourier space.

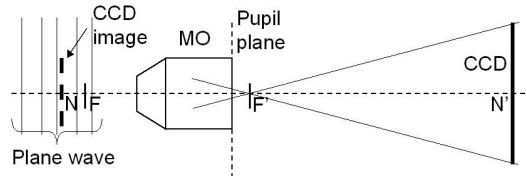

**Fig. 3** – Plane wave imaged through MO and transformed in a spherical wave.

Phase curvature introduced by the MO is depicted Fig. 3. Considering an object located in the CCD image plane and diffracting a plane wave, the wave impinging the CCD plane is a spherical wave emerging form F'. For most of the high-magnification MO, the back-focal plane F' and the pupil are the same. Thus, the phase curvature induced by the MO is equal to the curvature used for the pupil reconstruction (Eq. , and Fig. 2(b)). It should noted that these considerations remain valid if the reference wave is not a plane wave. In this case, only the reconstruction distance $z$ used to bring the pupil back to focus is modified.

For exact reconstruction in the object half space, the pixel pitches in the CCD image plane $\delta'_x$ have to be determined. Therefore, magnification of the imaging system $G = ON'/ON = \delta_x/\delta'_x$ has to be known. This can be done by comparing the size of the MO pupil with its holographic reconstruction. In the Fourier space, $\delta'_x$ is given by

$$\delta'_x = |F'N'|\delta_k / k$$

where $|F'N'|$ is CCD to pupil distance, and $\delta_k$ is the pixel pitch in the Fourier space. Considering the well-known FFT uncertainty principle ($N\delta'_x\delta_k = 2\pi$), and estimating $\delta'_x$ by comparing pupil size (in meters) and pupil reconstruction size (in pixels), one can obtain the CCD to pupil distance $|F'N'|$ using Eq.

$$|F'N'| = k\delta_x \delta'_x N / 2\pi$$

Knowing the focal distance F'O of the MO, the imaging system magnification becomes

$$G = |F'N'|/|F'O|$$

## 3. Image of the object reconstruction

Starting from the CCD image plane $H(x,y,z)$, image of the object $H(x,y,z')$ is recovered using angular spectrum propagation

$$H(x,y,z') = \mathsf{F}^{-1}\left\{\mathsf{F}\left[H(x,y,z)\right]\exp\left[-i(z'-z)\sqrt{k_n^2 - k_x^2 - k_y^2}\right]\right\}$$

Where $k_n = k/n$ is the wave vector taking the refractive index of the medium into account. This method is suited for short propagation distances $z - z'$ and operates with a constant pixel pitch, which is advantageous for three-dimensional distortion-free object reconstruction [6].

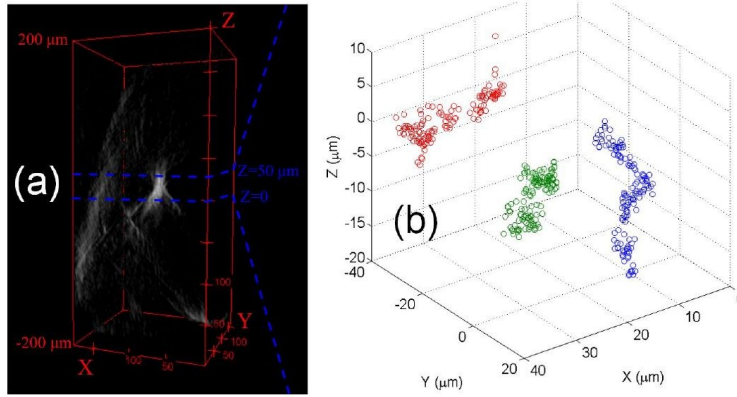

**Fig. 4** – (a) Three-dimensional reconstruction of the optical field intensity $|H(x; y; z')|^2$ diffracted by a 100 nm bead in agarose gel. *x; y* scales in pixels (160 nm), *z* scale in µm. (b) 3D trajectories of 3 particles of diameter of 100 nm in brownian motion in water (red, green, blue). Trajectories are reconstructed from 200 successive CCD frames.

## 4. Reconstruction examples

Figure 4 illustrates the three-dimensional reconstruction obtained with the proposed scheme. Here, the 3D optical field intensity $|H(x; y; z')|^2$ diffracted by a 100 nm bead in agarose gel calculated through Eq. is presented (Fig. 4(a)). Reconstruction distances ranges from -200 µm to 200 µm. Figure 4(b) shows the browian motion trajectories of three 100 nm in diameter particles in water. For each CCD frame, reconstruction is performed within a range of distances $z'$. Reconstruction distance is finally given by the maximum of $|H(x; y; z')|^2$.

## 5. Conclusion

We proposed a two-step distortion-free reconstruction method highly suitable for holographic microscopy. This method enables optimal filtering of the signal, and allows to compensate for phase distortion introduced by the MO. Experimental validation has been realized through the reconstruction of volumetric sample holograms.